# Approaches to threshold meson production [1]

Colin Wilkin

University College London, Gower St., London WC1E 6BT, UK

**Abstract**

Near-threshold data on meson production in nucleon-nucleon scattering are analysed in the short-range limit which permits a quantitative comparison of the production in two and three-body final states. The relative strengths of different meson productions are broadly in line with the predictions of one-meson-exchange models and it is deviations from these which are indications of extra physics.

Hans-Otto Meyer has given a very nice introduction to threshold meson production, with special emphasis on the pion. Though I shall draw on the experience gained in the pion case, the main thrust of my talk will be a description of the production of heavier non-strange mesons $X = \eta$, $\omega$, $\eta'$, $\phi$, leaving strangeness reactions to Wolfgang Eyrich. In my limited time I can only discuss nucleon-nucleon interactions and my theoretical analysis will be only a <u>zeroth</u> order approach to the understanding of these reactions. As Nathan Isgur put it in the conference introduction, this is mainly designed to promote understanding rather than provide a definitive calculation. Nevertheless, it might at least indicate the sensitivity of the results to the physics assumptions, and therefore what one might hope to learn from more refined models.

Most analyses of the $NN \to NNX$ reaction have been carried out in one-boson-exchange models, shown schematically in fig. 1, which serves to define the kinematics in the overall c.m. system. Letting $\vec{k}$ be the meson momentum, and $2\vec{q}$ the relative $NN$ momentum in the final state, then non-relativistically the c.m. kinetic energy $Q$ of the $NNX$ system is

$$Q = \frac{1}{2\mu_R}k^2 + \frac{1}{m}q^2 \; . \qquad (1)$$

Here $m$ is twice the nucleon reduced mass, $\mu$ the meson mass and $\mu_R$ the overall reduced mass equal to $\mu/(1 + \mu/2m)$. Data are often presented in terms of $\eta$, the maximum c.m. momentum of the meson in units of the meson mass

$$\eta = \sqrt{2\mu_R Q}/\mu \; . \qquad (2)$$

Calculations differ according to what mesons $x$ are exchanged in the diagram, whether distortion of the incident $NN$ waves or rescattering of the meson $x$ are included. However what is crucial in any description is a reasonable treatment of the nucleon-nucleon final-state interaction (FSI), drawn here as a blob. This is because of the nearby poles in the $S$-wave $NN$ amplitudes at $q^2 = -\alpha^2$ corresponding to the deuteron bound state in the $^3S_1$ channel or the $^1S_0$ virtual state. Taken together with the phase space factors, these poles tend to determine much of the energy dependence of the total cross section for meson production. Furthermore, in any region where these poles dominate, it is possible to link quantitatively meson production in cases where the two nucleons emerge separately or as a bound deuteron state.

---

[1] Invited talk at the Baryon98 conference, Bonn, September 1998

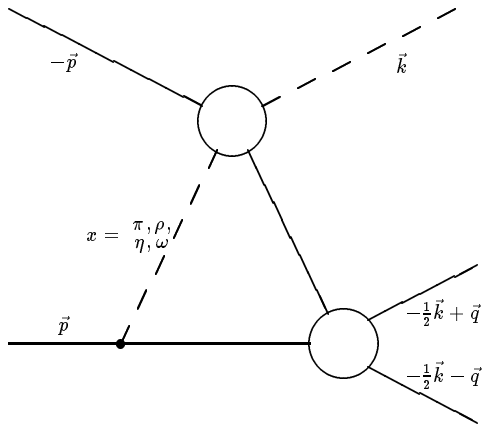

Figure 1: One-boson-exchange mechanism for meson production

An elegant way of deriving this latter relation can be found by working in $r$-space. Suppose $u(r)$ and $v(q,r)$, satisfying real boundary conditions, represent respectively the S-wave bound and scattering wave functions in the same local potential where, for large $r$, $u(r) \approx N\mathrm{e}^{-\alpha r}$. Then it can be shown that

$$v(q,r) \approx -\frac{1}{\sqrt{2\alpha(\alpha^2 + q^2)}}\, u(r) \,. \tag{3}$$

This result is <u>exact</u> when extrapolated to the pole at $q^2 = -\alpha^2$, but it is a remarkably robust approximation for a wide variety of potentials at small values of $r$ provided that $q^2$, $\alpha^2 \ll$ typical potential strength and $qR \ll 1$, $\alpha R \ll 1$, where $R$ is the typical potential range [1, 2]. In particular it works well for realistic $^3S_1$ $np$ potentials, despite ignoring the coupling to the $^3D_1$ channel [3].

The production of a pion or heavier meson in nucleon-nucleon scattering necessarily involves very large momentum transfers and so it is primarily the short-range part of the production operator which is tested in such processes. However, it is precisely at short distances that the extrapolation theorem allows us to approximate the scattering wave function in terms of that for the bound state. As a consequence, independent of the details of the operators, the production amplitudes $\mathcal{M}$ are linked by

$$\mathcal{M}(NN \to \{NN\}_q X) \approx -\mathcal{M}(NN \to \{NN\}_{bs} X)/\sqrt{2\alpha(q^2 + \alpha^2)} \,. \tag{4}$$

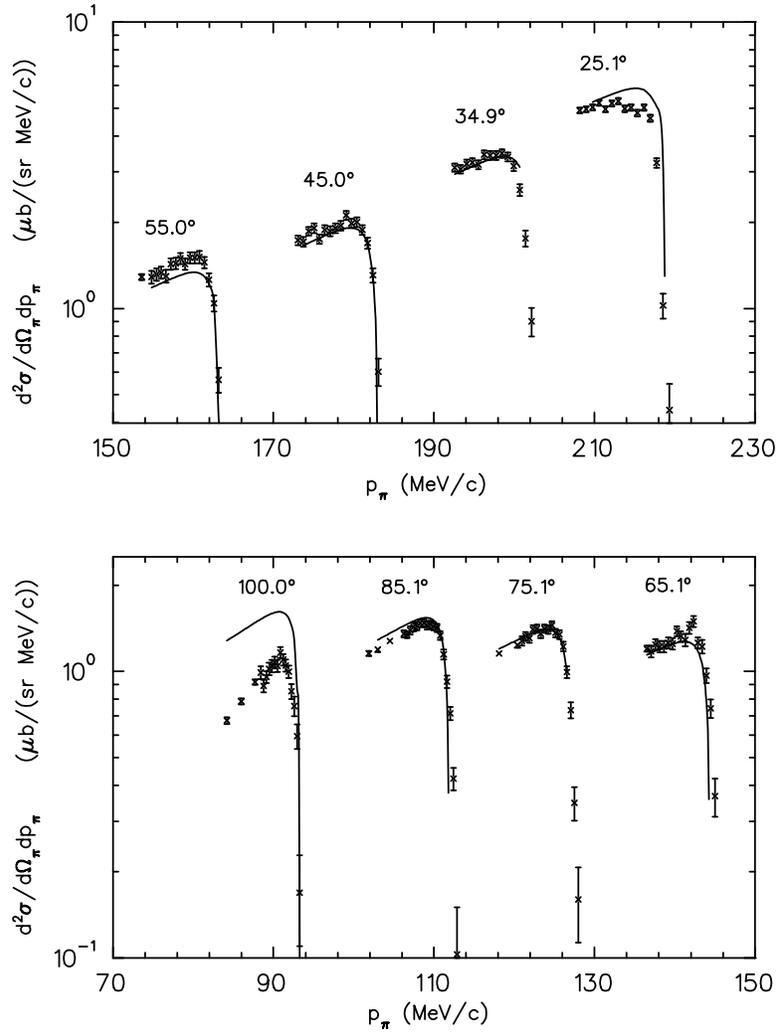

Figure 2: $pp \to pn\pi^+$ differential cross section at 420 MeV as a function of the pion angle and momentum compared to expectations on the basis of Eq. 5 transformed to the laboratory frame [4]. The upper end points correspond to $Q_{pn} = 0$ and the data are typically shown over a range of $\approx$20 MeV in excitation energy.

The above relation is of course only valid for $S$-wave spin-triplet $NN$ final states, where the bound state is the deuteron. However the wave function for the virtual state in the spin-singlet system with $\alpha < 0$ has an energy dependence dominated by a similar $1/\sqrt{q^2 + \alpha^2}$ factor to the bound state case of Eq. 4. The main difference between the two cases is that there is then no equivalent of the deuteron channel to normalise the cross section.

The cleanest place to test such an approach is in the comparison of the cross sections for $pp \to pn\pi^+$ and $pp \to d\pi^+$ away from threshold but in those parts of phase space where the $np$ excitation energy $Q_{pn} = q^2/m$ remains small. Under these conditions we expect that the final $S$-wave triplet contribution to the differential cross section should be

$$\frac{d^2\sigma}{d\Omega\,dx}(pp \to \{np\}\pi^+) \approx \frac{k(x)}{k(-1)} \frac{\sqrt{x}}{2\pi(x+1)} \frac{d\sigma}{d\Omega}(pp \to d\pi^+) \,. \quad (5)$$

The dimensionless variable $x$ is defined as $x = Q_{pn}/\epsilon = q^2/m\epsilon$, where $\epsilon$ is the deuteron binding energy, and $k(x)$ and $k(-1)$ are the momenta of the pion in the three and two-body reactions respectively.

The most detailed measurements of pion production have been carried out by a TRIUMF group [4] and the predictions of Eq. 5 are shown in fig. 2 for the preliminary data at 420 MeV. The normalisation of the predictions does not allow much room for the production of singlet $np$ final states, and this is in line with other information known on pion production. The proton analysing power $(A_y)$ predictions are equally successful [4].

Applying the formalism to near-threshold production, the condition on $Q_{np}$ is always met. If the energy dependence of the two-body cross section is of the form

$$\sigma_T(pp \to d\pi^+) = A\eta + B\eta^3 \,, \quad (6)$$

the integrals over phase space can be performed analytically [3] to give

$$\sigma_T(pp \to pn\pi^+) \approx \tfrac{1}{4}\zeta^3\eta^4 \left(1 + \sqrt{1 + \zeta^2\eta^2}\right)^{-2} \times$$
$$\left\{ A + \tfrac{1}{2}B\eta^2 \left[1 + \tfrac{1}{2}\eta^2\zeta^2 \left(1 + \sqrt{1 + \zeta^2\eta^2}\right)^{-2}\right] \right\} \,. \quad (7)$$

The only dependence upon the deuteron properties is through the parameter $\zeta = \mu_R/\sqrt{2\mu\epsilon}$. It should be mentioned that the approach has to be modified slightly to take into account, in an approximate way, external Coulomb corrections which are very important near threshold [5].

Data on total pion production cross sections are presented in fig. 3. By determining the parameters $A$ and $B$ from the $NN \to d\pi$ energy variation, a prediction for the triplet contribution to $pp \to pn\pi^+$ can be made. The small singlet contribution could be derived from the $pp \to pp\pi^0$ data using isospin arguments with Coulomb corrections, and it should be noted that the energy dependence of this cross section is similar to that of Eq. 7 with $B = 0$.

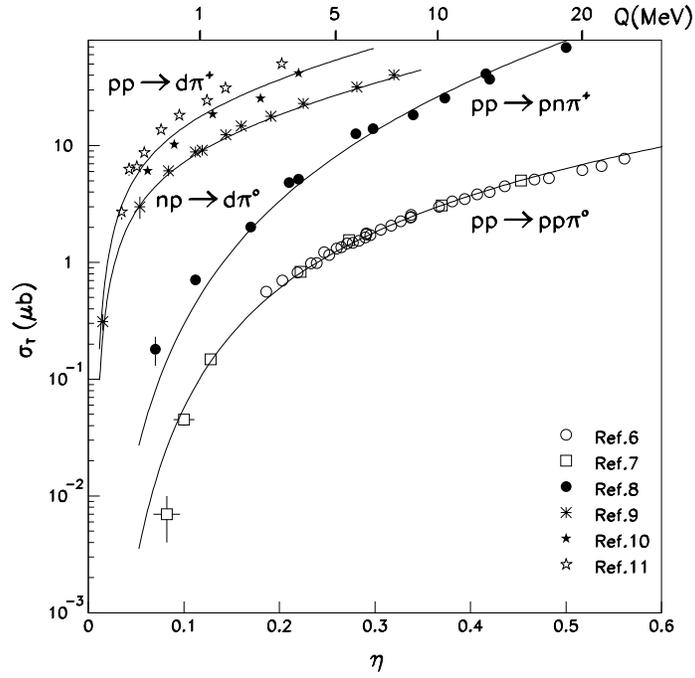

Figure 3: Prediction of Eq. 7 for the $pp \to pn\pi^+$ total cross section in terms of those for $pp \to d\pi^+(np \to d\pi^0)$ and $pp \to pp\pi^0$. Data are taken from Refs. [6, 7, 8, 9, 10, 11].

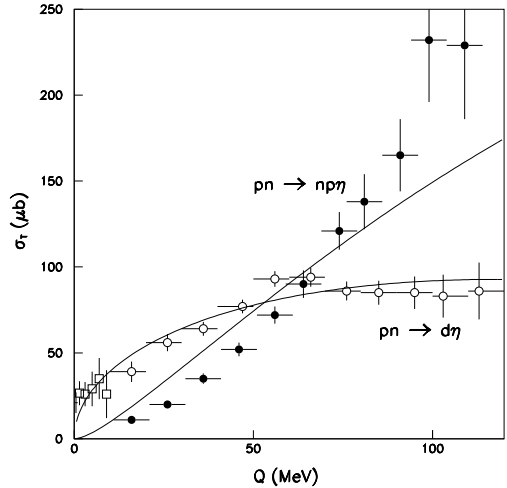

Figure 4: Prediction of Eq. 7 for the $pn \to pn\eta$ total cross section [13] in terms of those for $pn \to d\eta$ and $pp \to pp\eta$ [12].

The argument can be extended to describe the pion angular distribution in the c.m. system. At $\eta \approx 0.4$ the ratio of the $\theta_\pi = 0^0$ to $90^0$ cross section is determined to be $1.6 \pm 0.2$ [8], whereas the final state interaction approach described here leads to $1.8 \pm 0.2$. Agreement with the proton analysing power is at about the same level of precision.

A similar comparison can be made for $\eta$ production using the CELSIUS data obtained from quasi-free production on the neutron in the deuteron [12, 13]. Though qualitatively correct, the agreement in fig. 4 is not as good as for the pion case and it is tempting to attribute this to strong rescattering of the $\eta$ meson on the deuteron. For the $\eta\,^3$He and $\eta\,^4$He systems, such an FSI leads to the very marked threshold enhancements shown in fig. 5 for the $pd \to \,^3$He$\,\eta$ [14, 15] and $dd \to \,^4$He$\,\eta$ reactions [16, 17]. The squares of the spin-averaged amplitudes vary by factors of 2-3 over a few MeV in c.m. energy. Such behaviour must correspond to nearby poles of the production amplitudes. Whether these poles come sufficiently close to the physical domain to be considered as $\eta$-nucleus quasi-bound states or not will depend upon the number of nucleons involved and the strength of the $\eta$-nucleon interaction. It should however be noted that such poles are not mere displacements of the N*(1535), which dominates much of low energy $\eta$-nucleon(nucleus) physics, in the presence of other nucleons but rather extra singularities generated through the large $\eta$-nucleon scattering length [18]. The CELSIUS $pn \to d\eta$ data [12], as well as the earlier Saturne results [19], are certainly consistent with a large $\eta$-deuteron scattering length which may in fact be associated with a nearby virtual state. Preliminary results on $d(\gamma,\eta)X$ and ${}^4\text{He}(\gamma,\eta)X$ from the TAPS group working at MAMI, which were presented at this conference [20], show clear evidence for an $\eta$-deuteron threshold enhancement as well as a possible signal of the $\eta{}^3\text{He}/{}^3\text{H}$ enhancement being produced quasi-free in ${}^4\text{He}$.

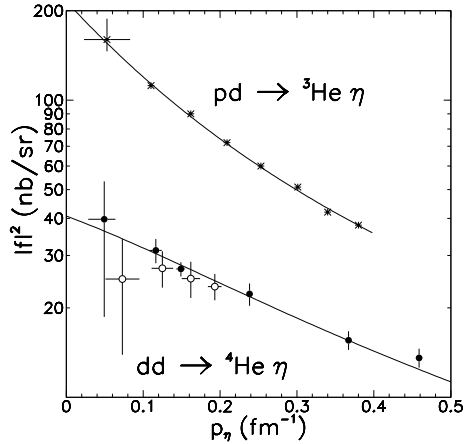

Figure 5: Averaged squared amplitudes for the $pd \to {}^3\text{He}\,\eta$ [15] and $dd \to {}^4\text{He}\,\eta$ reactions [16, 17] compared to the shapes obtained in a simultaneous optical potential fit [17] which indicates the existence of nearby quasi-bound states.

Turning now to proton-proton scattering, the short-range approximation to the pion-exchange diagram of fig. 1 predicts a total cross section of [21]

$$\sigma_T(pp \to ppX) = C\,\frac{(m+\mu)^2}{(2m+\mu)^{5/2}}\,\frac{\sqrt{\mu}}{(m\mu+m_\pi^2)^2}\,|f(\pi^0 p \to p\,X)|^2\left(\frac{Q}{1+\sqrt{1+Q/\epsilon}}\right)^2 \tag{8}$$

where, including Coulomb distortion, $\epsilon \approx 0.45$ MeV. In addition to the amplitude $f(\pi^0 p \to p\,X)$ for the production of meson $X$ in pion scattering, one also recognises the last term here as being the $S$-wave FSI factor of Eq. 7. The normalisation constant $C$ is close to that required to reproduce the $\eta$-production data [27], but if these data are rather used to determine the value of $C$, then this has to be multiplied by factors of 1.3 and 3 in order to describe well the $\omega$ and $\eta'$ data in fig. 6. It is therefore clear that there are no gross departures from the most naive implementation of a one-pion-exchange model. We are therefore going to have to look in much greater detail and at more exclusive observables in order to see features which are dependent upon the particular meson produced.

Simple isospin arguments, combined with the values of the $S$-wave $NN$ wave functions at short distances, suggest that the production of isoscalar mesons through one-pion exchange should be about four times higher in $pn$ collisions than $pp$ and the recent TSL measurement of $pn \to pn\eta$ shows a cross section with a similar energy dependence to that of $pp \to pp\eta$ but a factor of 6.5 times stronger [13]. This implies that some other exchange is also very important. This could be the $\rho$ meson since the photon has a significant coupling to the N*(1535) isobar. Denoting by $t_\pi$ the amplitude for $\pi$ exchange, and similarly for $\rho$, $\eta$ and $\omega$, the TSL data are consistent with [13]

$$\frac{\mid t_\pi - t_\rho + (t_\omega - t_\eta)/3 \mid}{\mid t_\pi + t_\rho + t_\omega + t_\eta) \mid} \approx 1.3, \qquad (9)$$

where the minus signs in the numerator arise from a combination of spin and isospin coupling.

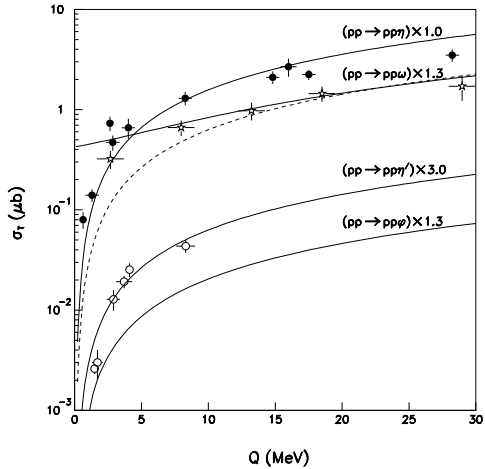

Figure 6: Total cross sections for meson production in the $pp \to ppX$ reaction near threshold. Experimental data on the production of $\eta$ [21, 22, 23], $\omega$ [24], and $\eta'$-mesons [21, 25] are compared with the predictions of Eq. 8 normalised to the $\eta$ data. The residual scale factors shown are close to unity except in the $\eta'$ case. The effect of the finite $\omega$ width on the predictions is shown in the change from the broken to the solid curve. The only data on $\varphi$ production is at $Q = 82$ MeV [26] and, whereas this is much too high for the $S$-wave assumptions to be valid, it is amusing that the curve shown for the $\phi$ does pass through this point.

Even if one neglects $\eta$ and $\omega$ exchange, there are still solutions where the $\rho$ dominates and others where it is the pion, and extra experimental information is required. This has been provided by a recent measurement of the $\eta$ angular distribution in $pp \to pp\eta$ at $Q = 37$ MeV [28]. In this reaction more $\eta$'s are produced at 90° than at 0° and this is in marked contrast with what is seen in $\pi^- p \to \eta n$ near threshold [29], but is in line with what is found in photoproduction [30]. Vector meson dominance then suggests strongly that $\rho$ exchange is the most important term in $\eta$ production in proton-proton collisions near threshold. The destructive interference between $\rho$ and $\pi$ exchange in the $pp$ case should enhance the effect and so we must expect the angular distribution in $pn \to d\eta$ to be flatter.

As stressed in the discussion by Norbert Kaiser, we should not be surprised if the production of $\eta$ and heavier mesons in nucleon-nucleon collisions is dominated by heavy meson exchange since the large momentum transfers involved picks out short-range physics. Such extra contributions might well account for the factor of three or more discrepancy for the $\eta'/\eta$ production ratio noted in fig. 6. However one should also note that any threshold enhancement present for the $\eta$ seems to be absent for the $\eta'$ since the near-threshold COSY-11 points [25], shown in fig. 6, seem rather to be suppressed near threshold as compared to the curve.

The predicted energy dependence of $\eta$ production shown in fig. 6 is not perfect, with the data points being a little high at low $Q$. This tendency is very similar to that noted for the $pn \to d\eta$ data in fig. 4 and might be due to strong $\eta$ rescattering in the final state generating a nearby virtual state.

When applying the formula of Eq. 8 to vector meson production, there is an uncertainty of a factor of 3/2 due to the ambiguity in the spin coupling but by far the biggest modification arises from the finite width of the meson which means that even at a nominal $Q = 0$ there is sufficient energy to produce the bottom half of the meson. After this effect is included, the energy dependence is well reproduced, apart possibly from the lowest point.

Experiments on $\omega$ production in the two-body reactions $\pi^- p \to n\omega$ [31] and $pd \to {}^3\text{He}\,\omega$ [32] have however quite unambiguously shown that, having taken the finite width effects very carefully into account, there is significant suppression in the production amplitudes for values of $Q$ which are a few times the meson width. This effect, which is illustrated in fig. 7 for the $^3$He case, might be associated with $\omega$ meson decay over the nucleon(ar) range and the subsequent rescattering of the decay pions. If there is a similar effect in $pp \to pp\omega$, it is clearly much smaller and this might be due to the data being averaged over the excitation energy in the final $pp$ system as well as over the $\omega$ width.

Comparison of $\omega$ and $\varphi$ production is interesting because of the interpretation of the ratio in terms of $(\omega, \varphi)$ mixing and the OZI rule. The lowest energy at which $pp \to pp\varphi$ has been measured corresponds to $Q \approx 82$ MeV [26], and this is far too high an energy for the $S$-wave assumptions used in Eq. 9 to be valid. Nevertheless it is amusing that the $\varphi$ prediction shown in fig. 6 does in fact pass through the preliminary experimental value. However one can only pass judgement on the validity of the OZI rule in the $pp \to pp\varphi/pp \to pp\omega$ ratio when one has dynamical models of both reactions, *i.e.* one needs calculations and not merely "understanding". A model of this kind, which has been produced by the Jülich group for the $\omega$ [33], is now being extended for the $\varphi$ [34].

It is of course vital to be able to describe the production of vector mesons in nucleon-nucleon collisions since there are hopes that the study of these mesons in a nuclear medium, possibly excited through heavy ion experiments, might shed light on the quark-gluon plasma or the restoration of chiral symmetry [35]. In particular it is important to know whether their production in proton-neutron collisions is much stronger than in proton-proton. To resolve this question, it is hoped to carry out measurements of quasi-free $pn \to d\,\omega(\varphi)$ at the COSY accelerator [36]. To go further one will then need data on angular distributions and spin observables and so there is work for years to come.

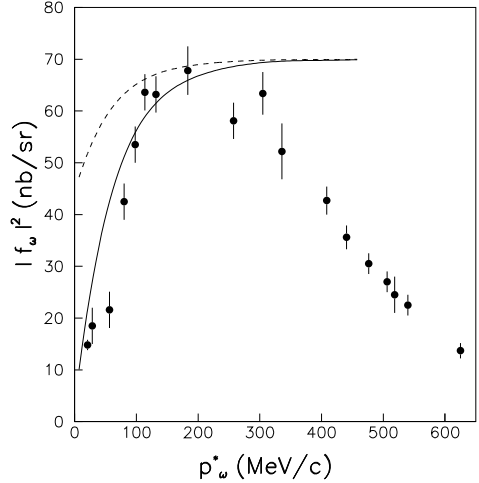

Figure 7: Average of the square of the amplitude for $\omega$ production in the $pd \to {}^3\text{He}\,\omega$ reaction showing the strong suppression near threshold [32]. Simulations obtained within the framework of a semi-classical rescattering of the decay pions from the $\omega$ decay by the final $^3$He nucleus are shown by the solid curve ($\pi^+\pi^-\pi^0$ decay) and dashed curve ($\pi^0\gamma$ decay) [32].